\documentclass[pra,aps,amsmath,amssymb,showkeys]{revtex4}
\usepackage{graphicx}
\usepackage[mathscr]{eucal}
\newcommand{\pen}{\openone}
\newcommand{\bro}{\boldsymbol{\rho}}

\newcommand{\me}{{\mathsf{E}}}
\newcommand{\mfe}{{\mathsf{F}}}
\newcommand{\cn}{{\mathbb{C}}}
\newcommand{\dn}{{\mathbb{D}}}

\newcommand{\rn}{{\mathbb{R}}}

\newcommand{\cald}{{\mathcal{D}}}
\newcommand{\calp}{{\mathcal{P}}}
\newcommand{\cle}{{\mathcal{E}}}
\newcommand{\clf}{{\mathcal{F}}}
\newcommand{\xdif}{{\mathrm{d}}}
\newcommand{\tr}{{\mathrm{tr}}}

\newcommand{\psymt}{{\mathsf{\Pi}}_{\mathrm{sym}}^{(t)}}
\newcommand{\psym}{{\mathsf{\Pi}}_{\mathrm{sym}}}

\begin{document}
\clearpage
\preprint{}

\title{R\'{e}nyi formulation of uncertainty relations for POVMs assigned to a quantum design}

\author{Alexey E. Rastegin}
\email{alexrastegin@mail.ru}
%\noaffiliation
\affiliation{Department of Theoretical Physics, Irkutsk State University, Irkutsk 664003, Russia}

\begin{abstract}
Information entropies provide powerful and flexible way to express restrictions
imposed by the uncertainty principle. This approach seems to be very
suitable in application to problems of quantum information theory.
It is typical that questions of such a kind involve measurements
having one or another specific structure. The latter often allows us
to improve entropic bounds that follow from uncertainty relations of
sufficiently general scope. Quantum designs have found use in many
issues of quantum information theory, whence uncertainty relations
for related measurements are of interest. In this paper, we obtain
uncertainty relations in terms of min-entropies and R\'{e}nyi
entropies for POVMs assigned to a quantum design. Relations of the
Landau--Pollak type are addressed as well. Using examples of quantum
designs in two dimensions, the obtained lower bounds are then
compared with the previous ones. An impact on entropic steering
inequalities is briefly discussed.
\end{abstract}


\maketitle

\pagenumbering{arabic}
\setcounter{page}{1}

\section{Introduction}\label{sec1}

The Heisenberg uncertainty principle \cite{heisenberg} is widely
recognized as a fundamental scientific concept. Since the first
formal derivations of Kennard \cite{kennard} and Robertson
\cite{robert} appeared, many approaches and scenarios were addressed
\cite{lahti,gour16}. In effect, the Heisenberg thought experiment
with microscope should be treated as dealing with successive
measurements. The scenario with successive measurements
\cite{mdsrin03,paban13} differs from the preparation one, when
repeated trials with the same quantum state are dealt with
\cite{rozp17}. An important question is how to characterize properly
the amount of uncertainties in quantum measurements. The usual way
in terms of lower bound on the product of variances has been
criticized for several reasons \cite{deutsch,maass}. As an
alternative, uncertainty bounds on the sum of variances were
examined \cite{huang12,macpa14}. Entropic uncertainty relations are
currently the subject of active researches
\cite{ww10,brud11,cbtw17,hall18,cerf19}. This approach allows us to
strengthen uncertainty relations due to quantum side information
\cite{bccrr10,cp2014,bww16} and connect them to fundamental properties of
relative entropies \cite{ccyz12}. Majorization technique is one of
powerful tools to formulate the uncertainty principle
\cite{fgg2013,prz2013,rpz2014,rz2016,zbig2018}. Another way to find
good entropic bounds is based on a direct optimization
\cite{zozor2013,zozor2014}. Uncertainty relations for successive
measurements have also been studied within the entropic approach
\cite{baek14,zzy15,rastann,rastent}.

Protocols of quantum information processing often deal with
measurements having some special inner structure. Mutually unbiased
bases are an especially important example reviewed, e.g., in the
paper \cite{bz10}. Symmetric informationally complete measurements
(SIC-POVMs) give another helpful tool for manipulating quantum
carriers of information. As the authors of \cite{rbksc04} showed,
the SIC-POVM problem is closely related to the concept of
$t$-designs. The existence of SIC-POVMs can be studied both
analytically and numerically, so that the list of solutions is
permanently growing \cite{scott10,scott17}. The claim that SICs
exist in every finite dimension is known as Zauner's conjecture in
its weakest form \cite{abfg19}. It is linked to a lot of purely
mathematical questions, some of them are discussed in
\cite{appleby05,fuchs17,af17,ab19}. Initially, spherical $t$-designs
on the unit sphere with some applications were studied in
\cite{delsart77}. Quantum $t$-designs also known as complex
projective designs have been examined for several reasons
\cite{scottjpa,ambain07}. In general, $t$-designs in
projective spaces were considered in \cite{hoggar82}. The concept of
designs is shown to be useful for a wide range of
information-theoretic applications
\cite{scottjpa8,scott9,dcel9,bhmn19,kwg19,cggz19}.

Due to an interesting structure of quantum designs and their
possible role in emerging technologies, we have come across several
questions. In particular, one aims to characterize the amount of
uncertainty in related quantum measurements. In comparison with the
known general formulations, more accurate estimates can be given for
quantum designs. Recently, the authors of \cite{guhne19} addressed
entropic uncertainty relations for POVMs assigned to a quantum
design. The used method is based on the monotonicity of certain
vector norms. It also allows one to write uncertainty relations in
terms of min-entropies. In general, these relations are not optimal.
As was discussed in \cite{rastmubs}, for measurements with a special
structure we can enhance an estimation of the corresponding
min-entropies. For MUBs and SIC-POVMs, this improvement holds due to
estimating indices of coincidence. It turns out that a similar
approach can be developed for quantum designs. Although formulation
becomes more complicated, the key idea is very similar to what was
exploited in \cite{rastmubs}. In effect, additional details are not
difficult from the viewpoint of calculations.

The aim of this work is to study R\'{e}nyi formulation of
uncertainty relations for POVMs assigned to a quantum design. The
proposed method mainly concentrates on a good estimation of
min-entropies. It is a natural development of the idea considered in
\cite{rastmubs}. Due to a non-obvious application of Jensen's
inequality, one can estimate the maximal probability from above.
This estimate also allows us to improve entropic uncertainty
relations for other values of the entropic parameter. The paper is
organized as follows. The preliminary material on quantum designs is
reviewed in Section \ref{sec2}. In particular, we summarize some
results on quantum designes and recall the used entropic functions.
Section \ref{sec3} is devoted to formulation of main results of the
paper. In Section \ref{sec4}, we consider examples of application to
concrete quantum designs as well as the comparison with the
previously given entropic bounds. An application to entropic
steering inequalities is briefly discussed. In Section \ref{sec5},
we conclude the paper with a summary of the results.

\section{Definitions and notation}\label{sec2}

In this section, we review the required material concerning quantum
designs and used entropies. Several equivalent ways to define
quantum designs were discussed in the literature. This concept can
be treated as an extension of the analog structure on spheres in
$\rn^{d}$. A spherical $t$-design is a finite set of normalized real
vectors such that the average value of any $t$th order polynomial
over this set is equal to the average over all normalized vectors in
$\rn^{d}$. The original motivation for studying these objects came
from the numerical evaluation of multi-dimensional integrals
\cite{cons98}.

In the finite-dimensional Hilbert space $\cn^{d}$, one considers lines passing through the origin, which
form the the complex projective space $\cn{P}^{d-1}$
\cite{scottjpa}. Up to a phase, each line can be represented by a
unit vector $|\phi\rangle$ in 
$\cn^{d}$. For a general discussion of complex projective spaces and
related topics, see chapter 4 of \cite{bengtsson}. The set
$\dn=\bigl\{|\phi_{k}\rangle:\,|\phi_{k}\rangle\in\cn^{d},\,\langle\phi_{k}|\phi_{k}\rangle=1,\,k=1,\ldots,K\bigr\}$
is a complex projective $t$-design, when the following property
holds \cite{scottjpa}. For every real polynomial $\calp_{t}$ of
degree at most $t$, the average value over $\dn$ is equal to the
average value over all normalized vectors of $\cn^{d}$, viz.
\begin{equation}
\frac{1}{K^{2}}\sum_{j,k=1}^{K}
\calp_{t}\Bigl(\bigl|\langle\phi_{j}|\phi_{k}\rangle\bigr|^{2}\Bigr)=
\int\!\int\xdif\mu(\psi)\,\xdif\mu(\psi^{\prime})\>
\calp_{t}\Bigl(\bigl|\langle\psi|\psi^{\prime}\rangle\bigr|^{2}\Bigr)
\, . \label{tdesdf}
\end{equation}
By $\mu(\psi)$, one denotes here the unique unitarily-invariant
probability measure on $\cn{P}^{d-1}$ induced by the Haar measure on
the corresponding unitary group. It is seen from the above
definition that each $t$-design is also a $s$-design with
$s\leq{t}$. It follows from the results of \cite{seym1984} that
$t$-designs in $\cn{P}^{d-1}$ exist for all $t$ and $d$. Examples
with smallest number of vectors are certainly difficult to
construct. Even if the smallest number of points is not required,
there is no general strategy to generate designs in all respective
cases. In more detail, these questions are discussed in
\cite{hardin96,gross07}. In effect, there exist important examples
that are widely used in applications. The concept of quantum designs
is naturally connected to the problem of building SIC-POVMs and
tight rank-one informationally complete POVMs
\cite{rbksc04,scottjpa}.

Complex projective designs have a lot of interesting properties. For
a $t$-design, one has \cite{scottjpa}
\begin{equation}
\frac{1}{K}\>\sum_{k=1}^{K}
|\phi_{k}\rangle\langle\phi_{k}|^{\otimes{t}}=\cald_{d}^{(t)}\,\psymt
\, , \label{topsy}
\end{equation}
where $\psymt$ is the projector onto the symmetric subspace of
$\bigl(\cn^{d}\bigr)^{\otimes{t}}$. The number $\cald_{d}^{(t)}$ is
the inverse of dimensionality of the symmetric subspace, namely
\begin{equation}
\cald_{d}^{(t)}=\binom{d+t-1}{t}^{\!-1}
=\frac{t!\,(d-1)!}{(d+t-1)!}
\ . \label{indim}
\end{equation}
The role of projector $\psymt$ was often emphasized in the
literature. In principle, the formula (\ref{topsy}) can be used as
equivalent definition of a quantum $t$-design \cite{scottjpa}. For
the given $t$, this formula can be applied for all positive integers
$s\leq{t}$. In particular, substituting $t=1$ results in
\begin{equation}
\frac{d}{K}\>\sum_{k=1}^{K}
|\phi_{k}\rangle\langle\phi_{k}|=\pen_{d}
\, . \label{comprel}
\end{equation}
Thus, unit vectors $|\phi_{k}\rangle$ lead to a resolution of the
identity in $\cn^{d}$. In principle, there may be several ways to
assign a set of POVMs to the given $t$-design. These ways are
unknown {\it a priori}, without an explicit consideration of kets
$|\phi_{k}\rangle$. The only obvious case is to take the complete
set $\cle$ consisting of operators
\begin{equation}
\me_{k}=\frac{d}{K}\>
|\phi_{k}\rangle\langle\phi_{k}|
\, . \label{mekdf}
\end{equation}
We shall also address the case, when $M$ rank-one POVMs
$\bigl\{\cle^{(m)}\bigr\}_{m=1}^{M}$ can be assigned to the given
quantum design. Each of these POVMs consist of $n$ operators of the
form
\begin{equation}
\me_{j}^{(m)}=\frac{d}{n}\>
|\phi_{j}^{(m)}\rangle\langle\phi_{j}^{(m)}|
\, . \label{mejdf}
\end{equation}
Here, the numbers $n$ and $M$ are connected by $K=nM$. In the
following, we will discuss an example of quantum design, in which
the kets form a set of mutually unbiased bases. When ways to choose
a set of POVMs are determined, for each of them our method will give
entropic uncertainty relations.

Let classical random variable $Z$ take values according to discrete
probability distribution $\{p(z)\}$. For $0<\alpha\neq1$, the
R\'{e}nyi $\alpha$-entropy is defined as \cite{renyi61}
\begin{equation}
R_{\alpha}(Z):=\frac{1}{1-\alpha}\>
\ln\!\left(\sum\nolimits_{z} p(z)^{\alpha}\right)
 . \label{repdf}
\end{equation}
This entropy does not increase with growth of $\alpha$. For a
discussion of basic properties of (\ref{repdf}), see section 2.7 of
\cite{bengtsson}. In the limit $\alpha\to1$, the right-hand side of
(\ref{repdf}) reduces to the Shannon entropy. In general,
information-theoretic functions of the R\'{e}nyi type do not succeed
all the properties of the standard functions. In more detail, these
questions are discussed in \cite{Kam98,ja04,rastkyb,rastrai}. In
particular, some popular measures to quantify mutual information are
not completely legitimate in the context of quantum cryptography
\cite{rastineq}. Nevertheless, the use of parametrized entropic
functions may often provide additional possibilities in analysis
\cite{maass}. A utility of generalized entropies in application to
combinatorial issues was shown in \cite{rastcomb}. The limit
$\alpha\to\infty$ gives the so-called min-entropy
\begin{equation}
R_{\infty}(Z)=-\ln\bigl(\max{p}(z)\bigr)
\, . \label{minen}
\end{equation}
Uncertainty relations in terms of min-entropies with some
application were discussed in \cite{MWB10,ngbw12}. In the following,
min-entropies will be used in posing uncertainty relations for
quantum designs.

If the pre-measurement state is described by density matrix $\bro$
of unit trace, then the probability of $j$th outcome is equal to
\begin{equation}
p_{j}(\cle^{(m)};\bro)=\frac{d}{n}\,\langle\phi_{j}^{(m)}|\bro|\phi_{j}^{(m)}\rangle
\, . \label{prokth}
\end{equation}
Substituting (\ref{prokth}) into (\ref{repdf}) leads to the entropy
$R_{\alpha}(\cle^{(m)};\bro)$. It follows from (\ref{topsy}) that,
for any density matrix $\bro$ and the given $t$-design, one has
\cite{guhne19}
\begin{equation}
\frac{1}{K}\>\sum_{k=1}^{K}
\langle\phi_{k}|\bro|\phi_{k}\rangle^{t}=\cald_{d}^{(t)}\,\tr\bigl(\bro^{\otimes{t}}\psymt\bigr)
\, . \label{indext}
\end{equation}
Combining (\ref{prokth}) with (\ref{indext}) then gives
\begin{equation}
\sum_{m=1}^{M}\sum_{j=1}^{n}p_{j}(\cle^{(m)};\bro)^{t}=
\left(\frac{d}{n}\right)^{\!t}\,\sum_{k=1}^{K}\langle\phi_{k}|\bro|\phi_{k}\rangle^{t}=
Kn^{-t}d^{\,t}\,\cald_{d}^{(t)}\,\tr\bigl(\bro^{\otimes{t}}\psymt\bigr)
\, . \label{mindexd}
\end{equation}
When a single POVM is assigned, we have $n=K$ and
\begin{equation}
\sum_{k=1}^{K}p_{k}(\cle;\bro)^{t}=
K^{1-t}d^{\,t}\,\cald_{d}^{(t)}\,\tr\bigl(\bro^{\otimes{t}}\psymt\bigr)
\, . \label{indexd}
\end{equation}
The formulas (\ref{mindexd}) and (\ref{indexd}) also holds with
$s=2,\ldots,t$ instead of $t$. The authors of \cite{guhne19,cirac18}
described how to express $\tr\bigl(\bro^{\otimes{t}}\psymt\bigr)$ as
a sum of monomials of the moments $\tr(\bro^{q})$. In particular, it
holds that
\begin{align}
\tr\bigl(\bro^{\otimes{2}}\psym^{(2)}\bigr)&=\frac{1}{2}
\left[1+\tr(\bro^{2})\right]
 , \label{ft2}\\
\tr\bigl(\bro^{\otimes{3}}\psym^{(3)}\bigr)&=\frac{1}{6}
\left[1+3\,\tr(\bro^{2})+2\,\tr(\bro^{3})\right]
 , \label{ft3}\\
\tr\bigl(\bro^{\otimes{4}}\psym^{(4)}\bigr)&=\frac{1}{24}
\left[1+6\,\tr(\bro^{2})+3\,\tr(\bro^{2})^{2}+8\,\tr(\bro^{3})+6\,\tr(\bro^{4})\right]
 . \label{ft4}
\end{align}
With growth of $t$, expressions of this kind become more
complicated. It is useful to note that
\begin{equation}
d^{\,t}\,\cald_{d}^{(t)}\,\tr\bigl(\bro_{*}^{\otimes{t}}\psymt\bigr)=1\leq{d}^{\,t}\,\cald_{d}^{(t)}\,\tr\bigl(\bro^{\otimes{t}}\psymt\bigr)
\, , \label{broas}
\end{equation}
where $\bro_{*}=d^{-1}\pen_{d}$ is the maximally mixed state.
Combining (\ref{indexd}) with (\ref{broas}) and
$\tr\bigl(\bro^{\otimes{t}}\psym\bigr)\leq1$ finally gives
\begin{equation}
K^{1-t}\leq\sum_{k=1}^{K}p_{k}(\cle;\bro)^{t}\leq
K^{1-t}d^{\,t}\,\cald_{d}^{(t)}
\, . \label{indexd1}
\end{equation}
The right inequality will be used to obtain state-independent
uncertainty relations. In the following, we will deal
with examples, where the right-hand side of (\ref{indexd1}) is
sufficiently small in comparison with $1$.

Finally, the main results of the paper \cite{guhne19} will be
recalled. To avoid bulky expressions, we introduce the two
parameters
\begin{align}
\bar{\beta}_{n}&=n^{1-t}d^{\,t}\,\cald_{d}^{(t)}\,\tr\bigl(\bro^{\otimes{t}}\psymt\bigr)
\, , \label{betan}\\
\bar{\beta}&=K^{1-t}d^{\,t}\,\cald_{d}^{(t)}\,\tr\bigl(\bro^{\otimes{t}}\psymt\bigr)
\, . \label{betak}
\end{align}
The latter follows from (\ref{betan}) by substituting $n=K$. Let $M$
rank-one POVMs $\cle^{(m)}$ be assigned to the given $t$-design. For $\alpha\geq{t}$,
the average $\alpha$-entropy satisfies \cite{guhne19}
\begin{equation}
\frac{1}{M}\>\sum_{m=1}^{M}R_{\alpha}(\cle^{(m)};\bro)\geq
\frac{\alpha}{t(1-\alpha)}\,\ln\bar{\beta}_{n}
\, . \label{gunaln}
\end{equation}
For the case of a single POVM, the above uncertainty relation
reduces to
\begin{equation}
R_{\alpha}(\cle;\bro)\geq\frac{\alpha}{t(1-\alpha)}\,\ln\bar{\beta}
\, , \label{gunal}
\end{equation}
These results are derived from (\ref{mindexd}) and (\ref{indexd})
due to monotonicity of the vector $p$-norm and the function
$y\mapsto-\ln{y}$. The inequalities (\ref{gunaln}) and (\ref{gunal})
remain valid for  $s$ instead of $t$ and $\alpha\geq{s}$, provided
that integer $s$ lies between $2$ and $t$. According to
(\ref{gunaln}) and (\ref{gunal}), the min-entropies obey
\begin{align}
\frac{1}{M}\>\sum_{m=1}^{M}R_{\infty}(\cle^{(m)};\bro)&\geq-\frac{1}{t}\,\ln\bar{\beta}_{n}
\, , \label{guninn}\\
R_{\infty}(\cle;\bro)&\geq-\frac{1}{t}\,\ln\bar{\beta}
\, . \label{gunin}
\end{align}
As will be shown, the uncertainty relations
(\ref{gunaln})--(\ref{gunin}) can be improved. The authors of
\cite{guhne19} also formulated uncertainty relations in terms of
Tsallis entropies. Such entropies are not considered in the
following.

\section{Main results}\label{sec3}

This section is devoted to deriving R\'{e}nyi-entropy uncertainty
relations for POVMs assigned to a quantum design. We begin with a
discussion of min-entropy uncertainty relations. To obtain
uncertainty relations from (\ref{mindexd}) and
(\ref{indexd}), we will use an auxiliary consideration. Let $n$
positive numbers $y_{j}$ obey the two relations
\begin{align}
\sum\nolimits_{j=1}^{n} y_{j}&=1
\, , \label{sum1}\\
\sum\nolimits_{j=1}^{n} y_{j}^{\,t}&=\beta
\, . \label{sumbt}
\end{align}
It follows from the normalization (\ref{sum1}) that
$n^{1-t}\leq\beta\leq1$. We aim to estimate maximum of the numbers
$y_{j}$ from above. For definiteness, we assume that these numbers
are arranged in non-decreasing order, so that $y_{j}\leq{y}_{n}$. By
convexity of the function $y\mapsto{y}^{t}$ for $t>1$, one has
\begin{equation}
\biggl(\frac{1}{n-1}\>\sum\nolimits_{j=1}^{n-1} y_{j}\biggr)^{t}
\leq\frac{1}{n-1}\>\sum\nolimits_{j=1}^{n-1} y_{j}^{\,t}
\, . \label{sumcon}
\end{equation}
Combining the left- and right-hand sides of (\ref{sumcon})
respectively with (\ref{sum1}) and with (\ref{sumbt}) finally gives
\begin{equation}
\frac{(1-y_{n})^{\,t}}{(n-1)^{t-1}}
\leq\beta-y_{n}^{\,t}
\, . \label{sumfon}
\end{equation}
To estimate $y_{n}$ from above, one should solve
\begin{equation}
\frac{(1-y)^{\,t}}{(n-1)^{t-1}\beta}+\frac{y^{\,t}}{\beta}=1
\, . \label{curkev}
\end{equation}
Restricting a consideration to the quadrant $I$, we wish to find
coordinates of the intersection of the curve
\begin{equation}
\frac{x^{\,t}}{(n-1)^{t-1}\beta}+\frac{y^{\,t}}{\beta}=1
\label{curvek}
\end{equation}
with the straightforward line $x+y=1$. The curve intersects the axes
with the abscissa $(n-1)^{1-1/t}\beta^{\,1/t}$ and with the ordinate
$\beta^{\,1/t}$. For even $t$, the equation (\ref{curvek}) gives an
oval line symmetric with respect to both the axes. It bounds the
convex set on the plane. The points of this oval line lie between
the ellipse with the above semi-axes and the corresponding
rectangle, closely to the latter. Here, we have two intersection
points or one point of touching the line $x+y=1$. With growth of
even $t$, the curve (\ref{curvek}) will mimic a rectangle with
rounded corners. For odd $t$, the curve goes inside the mentioned
rectangle only in the quadrant $I$. Beyond this quadrant, the curve
tends to go along the asymptote
\begin{equation}
y=-\,\frac{x}{(n-1)^{1-1/t}}
\ . \label{asymp}
\end{equation}
When $n>2$, the latter intersects $x+y=1$ in the quadrant $IV$.
Hence, we conclude that there is two or three intersection points.
In all the cases, we are interested in the point with maximal
ordinate.

By $\Upsilon_{n-1}^{(t)}(\beta)$, we further denote the maximal real
root of (\ref{curkev}). Some properties of
$\Upsilon_{n-1}^{(t)}(\beta)$ as a function of $\beta$ are discussed
in Appendix \ref{roots}. Let us mention a special case to be
discussed explicitly. For $\beta=n^{1-t}$, the equation
(\ref{curkev}) has the root $y_{n}=1/n$, so that
\begin{equation}
\Upsilon_{n-1}^{(t)}\bigl(n^{1-t}\bigr)=\frac{1}{n}
\ . \label{upcom}
\end{equation}
Writing the latter as $\beta^{1/(t-1)}$ with $t\geq2$, one can
expect concavity and increasing, at least in some neighborhood. The
answer (\ref{upcom}) is naturally explained as follows. For the
maximally mixed state, we have
$p_{j}(\cle^{(m)};\bro_{*})=1/n$ irrespectively to $j$. Hence, the
maximal probability is given by (\ref{upcom}). For $n=K$, the
left-hand side of (\ref{indexd1}) is also reached with the
maximally mixed state.

For $t\geq5$, one is generally unable to express
$\Upsilon_{n-1}^{(t)}(\beta)$ analytically using radicals. On the
other hand, for the given parameters the answer can always be found
by appropriate numerical procedure with any desired accuracy. In
the cases $t=2,3,4$, we can express the answer in a closed analytic
form. The case $t=2$ is the simplest one, when
\begin{equation}
\Upsilon_{n-1}^{(2)}(\beta)=\frac{1}{n}
\left(1+\sqrt{n-1}\sqrt{n\beta-1}\,\right)
 . \label{upst2}
\end{equation}
This result was derived and applied to uncertainty relations in the
paper \cite{rastmubs}. Some useful formulas for $t=3$ are given in
Appendix \ref{cas34}. Due to the above consideration, the following
statement takes place.

\newtheorem{pp1}{Proposition}
\begin{pp1}\label{res1}
Let $M$ rank-one POVMs $\cle^{(m)}$, each with $n$ elements of the
form (\ref{mejdf}), be assigned to a quantum $t$-design
$\dn=\bigl\{|\phi_{k}\rangle\bigr\}_{k=1}^{K}$ in $d$ dimensions. If
the pre-measurement state is described by the density matrix $\bro$,
then
\begin{equation}
\frac{1}{M}\>\sum_{m=1}^{M}R_{\infty}(\cle^{(m)};\bro)\geq
-\ln\bigl(\Upsilon_{n-1}^{(t)}(\bar{\beta}_{n})\bigr)
\, , \label{eqren1}
\end{equation}
where $\Upsilon_{n-1}^{(t)}(\beta)$ denotes the maximal real root of
(\ref{curkev}) and $\bar{\beta}_{n}$ is defined by (\ref{betan}).
\end{pp1}

{\bf Proof.} It follows from the preliminary consideration that, for
all $m=1,\ldots,M$,
\begin{equation}
\underset{j}{\max}\,p_{j}(\cle^{(m)};\bro)\leq\Upsilon_{n-1}^{(t)}(\beta_{m})
\, , \label{maxjb}
\end{equation}
where
\begin{equation}
\beta_{m}=\sum_{j=1}^{n}p_{j}(\cle^{(m)};\bro)^{t}
\, . \nonumber
\end{equation}
Combining (\ref{minen}) with (\ref{maxjb}) then gives
$R_{\infty}(\cle^{(m)};\bro)\geq-\ln\bigl(\Upsilon_{n-1}^{(t)}(\beta_{m})\bigr)$,
whence we write
\begin{align}
\frac{1}{M}\>\sum_{m=1}^{M}R_{\infty}(\cle^{(m)};\bro)\geq
\sum_{m=1}^{M}\frac{1}{M}\,\bigl[{}-\ln\bigl(\Upsilon_{n-1}^{(t)}(\beta_{m})\bigr)\bigr]
&\geq-\ln\!\left(\,\sum_{m=1}^{M}\frac{1}{M}\,\Upsilon_{n-1}^{(t)}(\beta_{m})\right)
\label{teps1}\\
&\geq-\ln\Upsilon_{n-1}^{(t)}\!\left(\,\sum\nolimits_{m=1}^{M}\frac{\beta_{m}}{M}\right)
 . \label{teps2}
\end{align}
The steps (\ref{teps1}) and (\ref{teps2}) respectively hold due to
convexity and decreasing of the function $y\mapsto-\,\ln{y}$, with
adding
\begin{equation}
\sum_{m=1}^{M}\frac{1}{M}\,\Upsilon_{n-1}^{(t)}(\beta_{m})\leq
\Upsilon_{n-1}^{(t)}\!\left(\,\sum\nolimits_{m=1}^{M}\frac{\beta_{m}}{M}\right)
 . \nonumber
\end{equation}
The latter is valid due to concavity of
$\Upsilon_{n-1}^{(t)}(\beta)$ with respect to $\beta$ (this fact is
shown in Appendix \ref{roots}). Combining (\ref{teps2}) with
(\ref{mindexd}) and $K=nM$ completes the proof of (\ref{eqren1}).
$\blacksquare$

When single POVM $\cle$ with $K$ elements (\ref{mekdf}) is assigned
to the given $t$-design, the inequality (\ref{eqren1}) reduces to
\begin{equation}
R_{\infty}(\cle;\bro)\geq-\ln\bigl(\Upsilon_{K-1}^{(t)}(\bar{\beta})\bigr)
\, , \label{eqres1}
\end{equation}
where $\bar{\beta}$ is defined by (\ref{betak}). The results
(\ref{eqren1}) and (\ref{eqres1}) provide state-dependent
uncertainty relations in terms of min-entropies. In particular, the
second one is expressed in terms of the single parameter
(\ref{betak}). The result (\ref{eqres1}) cannot further be improved
without using additional data about the actual pre-measurement
state. As the function $y\mapsto-\ln{y}$ decreases, we also have
\begin{align}
\frac{1}{M}\>\sum_{m=1}^{M}R_{\infty}(\cle^{(m)};\bro)
&\geq-\ln\bigl(\Upsilon_{n-1}^{(t)}(n^{1-t}d^{\,t}\,\cald_{d}^{(t)})\bigr)
\, , \label{eqres1sn}\\
R_{\infty}(\cle;\bro)&\geq-\ln\bigl(\Upsilon_{K-1}^{(t)}(K^{1-t}d^{\,t}\,\cald_{d}^{(t)})\bigr)
\, . \label{eqres1si}
\end{align}
These state-independent formulations hold for all states and
correspond to substituting the right-hand side of (\ref{indexd1}).

The presented method immediately leads to uncertainty relations of
the Landau--Pollak type. The original results of Landau and Pollak
concern uncertainty in signal theory \cite{pollak61}. The authors of
\cite{maass} gave reformulation to characterize the amount of
uncertainty in projective quantum measurements. Extensions to POVM
measurements were formulated in \cite{immy2007,bosyk2014}. In
contrast to projective measurements, for a single rank-one POVM we may have a
non-trivial upper bound on the maximal probability. This takes
place, when number of outcomes exceeds the dimensionality. Due to
(\ref{maxjb}) and concavity of $\Upsilon_{n-1}^{(t)}(\beta)$ with
respect to $\beta$, we obtain
\begin{align}
\frac{1}{M}\>\sum_{m=1}^{M}\underset{j}{\max}\,p_{j}(\cle^{(m)};\bro)&\leq
\sum_{m=1}^{M}\frac{1}{M}\,\Upsilon_{n-1}^{(t)}(\beta_{m})
\leq\Upsilon_{n-1}^{(t)}(\bar{\beta}_{n})
\, , \label{lpmax}\\
\underset{k}{\max}\,p_{k}(\cle;\bro)
&\leq\Upsilon_{K-1}^{(t)}(\bar{\beta})
\, , \label{lpmak}
\end{align}
These relations can be used in formulating criteria to characterize
entanglement or steerability. For instance, the writers of
\cite{vicen05} considered separability conditions based on the
Landau--Pollak uncertainty relation.

The Newton--Raphson method is a well-known numerical algorithm for
finding roots of equations. In application to $f(y)=0$ with some
initial guess $y^{(0)}$, this method gives a correction
\begin{equation}
y^{(1)}-y^{(0)}=-\,\frac{f(y^{(0)})}{f^{\prime}(y^{(0)})}
\ . \label{nr1st}
\end{equation}
The value $y^{(1)}$ is a better approximation of the root than
$y^{(0)}$. By repeating such steps, one is able to improve
approximations successively. In order to calculate
$\Upsilon_{n-1}^{(t)}(\bar{\beta}_{n})$, we can start the process
with the value $\bar{\beta}_{n}^{\,1/t}$. The latter is larger than
the desired root, since the condition (\ref{sumbt}) for
$\beta=\bar{\beta}_{n}$ implies
\begin{equation}
y_{j}^{\,t}\leq\bar{\beta}_{n}
\, . \nonumber
\end{equation}
In the case of interest, we use the function
$y\mapsto(n-1)^{t-1}y^{t}+(1-y)^{t}-(n-1)^{t-1}\bar{\beta}_{n}$. The
latter is obviously convex and increasing for $y>1/n$. Substituting
$\bar{\beta}_{n}^{\,1/t}$ then implies a positive value of the
function, which should vanish for the desired point. Since the
Newton--Raphson method replaces the function with its tangent line,
the first step will result in the term that exceeds the root due to
convexity. In other words, we have
\begin{equation}
\Upsilon_{n-1}^{(t)}(\bar{\beta}_{n})\leq\widetilde{\Upsilon}_{n-1}^{(t)}(\bar{\beta}_{n})=\bar{\beta}_{n}^{\,1/t}-
\frac{\bigl(1-\bar{\beta}_{n}^{\,1/t}\bigr)^{t}}{t(n-1)^{t-1}\bar{\beta}_{n}^{\,1-1/t}-t\bigl(1-\bar{\beta}_{n}^{\,1/t}\bigr)^{t-1}}
\ . \label{upsst1}
\end{equation}
This explicit expression is slightly complicated, but quite suitable
to calculate. Combining (\ref{eqren1}) and (\ref{eqres1}) with
(\ref{upsst1}), one gets
\begin{align}
\frac{1}{M}\>\sum_{m=1}^{M}R_{\infty}(\cle^{(m)};\bro)
&\geq-\ln\bigl(\widetilde{\Upsilon}_{n-1}^{(t)}(\bar{\beta}_{n})\bigr)
\, , \label{eqres11n}\\
R_{\infty}(\cle;\bro)&\geq-\ln\bigl(\widetilde{\Upsilon}_{K-1}^{(t)}(\bar{\beta})\bigr)
\, . \label{eqres11}
\end{align}
That is, one Newton--Raphson step gives valid inequalities whose
right-hand sides can be expressed analytically. Due to positivity
and convexity, repeated steps of such a kind will further improve
our result. However, the corresponding expressions are sufficiently
bulky.

Due to (\ref{eqren1}), we can also estimate the average
$\alpha$-entropy from below for all $\alpha\geq{t}$. This is
obtained by some extension of the reasons proposed in
\cite{rastosid}. For $\alpha\geq{t}$, we merely write
\begin{equation}
\sum\nolimits_{z} p(z)^{\alpha}\leq
\bigl(\max{p}(z)\bigr)^{\alpha-t}\,\sum\nolimits_{z} p(z)^{t}
\, . \label{prares2}
\end{equation}
The latter together with (\ref{repdf}) implies
\begin{equation}
R_{\alpha}(Z)\geq
\frac{\alpha-t}{\alpha-1}\,R_{\infty}(Z)+
\frac{t-1}{\alpha-1}\,R_{t}(Z)
\, , \label{alteq}
\end{equation}
whenever $\alpha\geq{t}\geq1$. It follows from (\ref{gunaln}) that
\begin{equation}
\frac{1}{M}\>\sum_{m=1}^{M}R_{t}(\cle^{(m)};\bro)\geq
\!{}-\frac{\ln\bar{\beta}_{n}}{t-1}
\ . \label{pemt}
\end{equation}
Combining (\ref{eqren1}), (\ref{alteq}) and (\ref{pemt}), we have
arrived at a conclusion.

\newtheorem{pp2}[pp1]{Proposition}
\begin{pp2}\label{res2}
Let $M$ rank-one POVMs $\cle^{(m)}$, each with $n$ elements of the
form (\ref{mejdf}), be assigned to a quantum $t$-design
$\dn=\bigl\{|\phi_{k}\rangle\bigr\}_{k=1}^{K}$ in $d$ dimensions.
For $\alpha\geq{t}$, it holds that
\begin{equation}
\frac{1}{M}\>\sum_{m=1}^{M}R_{\alpha}(\cle^{(m)};\bro)
\geq-\,\frac{\alpha-t}{\alpha-1}\,\ln\bigl(\Upsilon_{n-1}^{(t)}(\bar{\beta}_{n})\bigr)
-\frac{\ln\bar{\beta}_{n}}{\alpha-1}
\ , \label{eqres2n}
\end{equation}
where $\Upsilon_{n-1}^{(t)}(\beta)$ denotes the maximal real root of
(\ref{curkev}) and $\bar{\beta}_{n}$ is defined by (\ref{betan}).
\end{pp2}

When single POVM $\cle$ with $K$ elements (\ref{mekdf}) is assigned
to the given $t$-design, the inequality (\ref{eqres2n}) reduces to
\begin{equation}
R_{\alpha}(\cle;\bro)\geq-\,\frac{\alpha-t}{\alpha-1}\,\ln\bigl(\Upsilon_{K-1}^{(t)}(\bar{\beta})\bigr)
-\frac{\ln\bar{\beta}}{\alpha-1}
\ , \label{eqres2}
\end{equation}
where $\bar{\beta}$ is defined by (\ref{betak}). Substituting
$\bar{\beta}_{n}=n^{1-t}d^{\,t}\,\cald_{d}^{(t)}$ into
(\ref{eqres2n}) leads to the state-independent formulation
\begin{equation}
\frac{1}{M}\>\sum_{m=1}^{M}R_{\alpha}(\cle^{(m)};\bro)
\geq-\,\frac{\alpha-t}{\alpha-1}\,\ln\bigl(\Upsilon_{n-1}^{(t)}(n^{1-t}d^{\,t}\,\cald_{d}^{(t)})\bigr)
-\frac{\ln(n^{1-t}d^{\,t}\,\cald_{d}^{(t)})}{\alpha-1}
\ . \label{eqres2in}
\end{equation}
Let us compare new entropic bounds with the previous ones. For
brevity, we focus on the results (\ref{gunal}) and (\ref{eqres2}).
It is instructive to apply them to the maximally mixed state
$\bro_{*}$. Using $\bar{\beta}=K^{1-t}$ and (\ref{upcom}), one gets
\begin{equation}
R_{\alpha}(\cle;\bro_{*})\geq\frac{\alpha-t}{\alpha-1}\,\ln{K}+\frac{t-1}{\alpha-1}\,\ln{K}=\ln{K}
\, . \label{upcom1}
\end{equation}
In other words, the uncertainty relation (\ref{eqres2}) is saturated
with the maximally mixed state. At the same time, the relation
(\ref{gunal}) reads here as
\begin{equation}
R_{\alpha}(\cle;\bro_{*})\geq\frac{\alpha(t-1)}{t(\alpha-1)}\,\ln{K}
\, . \label{gunalcom}
\end{equation}
Restricting to $\alpha\geq{t}$, the latter coincides with
(\ref{upcom1}) only for $\alpha=t$. For sufficiently large $\alpha$,
the difference between these bounds is approximately $\ln{K}/t$.
Thus, the result (\ref{eqres2}) sometimes provides a considerable
improvement of (\ref{gunal}). Taking $\alpha=\infty$, the ratio of
the right-hand sides of (\ref{upcom1}) and (\ref{gunalcom}) is
$t/(t-1)$. The latter implies $1.5$ for $3$-designs and $1.25$ for
$5$-designs. For other states, when $\bar{\beta}>K^{1-t}$, the
amount of improvement is lesser. Nevertheless, the inequality
(\ref{eqres2}) is stronger than (\ref{gunal}). In a similar manner,
the result (\ref{eqres2n}) somehow enhances (\ref{gunaln}).

\section{Examples of uncertainty bounds for quantum designs}\label{sec4}

In this section, we consider examples of application of the
developed method to concrete quantum designs in two dimensions. The
short description of these designs in terms of components of the
Bloch vector can be found in \cite{guhne19}. The
corresponding vertices form some polyhedron. We will mainly focus on
the case of single assigned POVM. It is instructive to visualize
distinctions between the lower estimates (\ref{gunin}),
(\ref{eqres1}), and (\ref{eqres11}). Note that the right-hand side of
(\ref{eqres11}) can be treated as an improvement of (\ref{gunin})
within the first Newton--Raphson step. In general, the significance
of correction depends on the quantity (\ref{betak}).

\begin{figure*}
\centering \includegraphics[height=7.6cm]{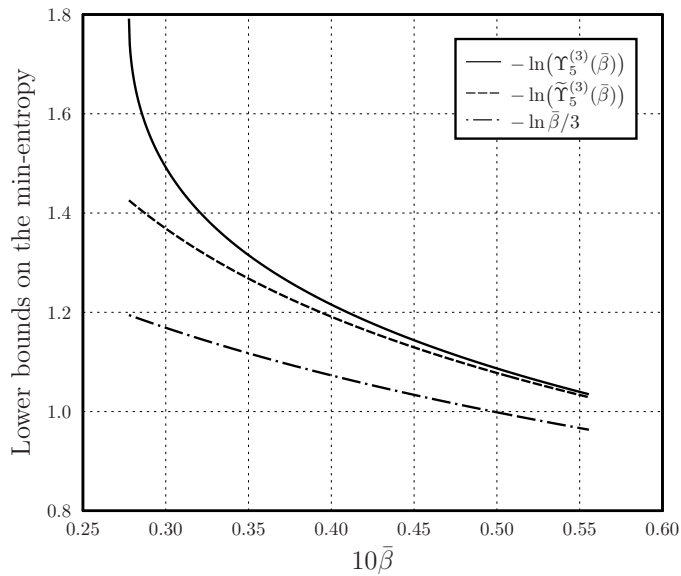}
\caption{\label{fig1} The three lower bounds on the min-entropy
versus the parameter (\ref{betak}) for the $3$-design. For
convenience of presentation, the abscissa shows $10\bar{\beta}$.}
\end{figure*}

Let us begin with the $3$-design with $K=6$ vertices forming an
octahedron. The quantity $\Upsilon_{5}^{(3)}(\bar{\beta})$ is
calculated in line with the formulas listed in Appendix \ref{cas34}.
In Fig. \ref{fig1}, we plot the lower bounds (\ref{gunin}),
(\ref{eqres1}), and (\ref{eqres11}). In general, the improvement due
to both the formulas (\ref{eqres1}) and (\ref{eqres11}) is
considerable. The right-hand sides of (\ref{eqres1}) and
(\ref{eqres11}) are close for pure states and states with a low
mixedness. Here, these results enhance the right-hand side of
(\ref{gunin}) by values of order $7$ \%. For small
values of $\bar{\beta}$, when the measured state is close to the
maximally mixed one, we see essential distinctions between all the
three uncertainty bounds. Due to (\ref{eqres1}), we improve the
lower entropic bound in comparison with (\ref{gunin}) approximately
$1.5$ times. Overall, the results (\ref{eqres1}) and (\ref{eqres11})
provide a good improvement.

\begin{figure*}
\centering \includegraphics[height=7.6cm]{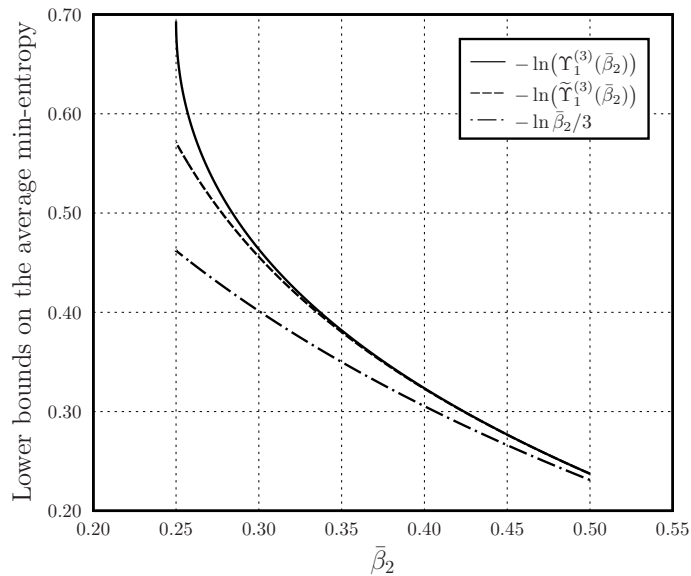}
\caption{\label{fig2} The three lower bounds on the average
min-entropy for the three mutually unbiased bases in two
dimensions.}
\end{figure*}

The example with octahedron is interesting in additional respect due
to the following. The considered $3$-design is formed by eigenstates
of the Pauli matrices. That is, the complete set of three mutually
unbiased bases are assigned to this $3$-design. Entropic uncertainty
relations for such bases have attracted a lot of attention. The
three bases associated with Pauli's matrices are traditionally used
as a test. Restricting a consideration to min-entropies, we shall
discuss them in more detail. Substituting $M=3$ and $n=2$, the
formula (\ref{eqren1}) reads as
\begin{equation}
\frac{1}{3}\>\sum_{m=1}^{3}R_{\infty}(\cle^{(m)};\bro)\geq
-\ln\bigl(\Upsilon_{1}^{(3)}(\bar{\beta}_{2})\bigr)=
-\ln\!\left(\frac{\sqrt{3}+\bigl(4\bar{\beta}_{2}-1\bigr)^{1/2}}{2\sqrt{3}}\right)
 , \label{eqren1mub}
\end{equation}
where we used (\ref{upsn2}). One has
$\bar{\beta}_{2}=\bigl[1+3\,\tr(\bro^{2})+2\,\tr(\bro^{3})\bigr]/12$
due to (\ref{ft3}) and (\ref{betan}). The relation (\ref{eqren1mub})
should be compared with the inequality
\begin{equation}
\frac{1}{3}\>\sum_{m=1}^{3}R_{\infty}(\cle^{(m)};\bro)\geq
-\,\frac{\ln\bar{\beta}_{2}}{3}
\ , \label{guninub}
\end{equation}
which follows from (\ref{guninn}). Another uncertainty relation in
terms of the purity $\tr(\bro^{2})$ was proved in \cite{rastmubs}.
For $M=3$ and $d=2$, it reads as
\begin{equation}
\frac{1}{3}\>\sum_{m=1}^{3}R_{\infty}(\cle^{(m)};\bro)\geq
\ln\!\left(\frac{2\sqrt{3}}{\sqrt{3}+\sqrt{2\,\tr(\bro^{2})-1}}\right)
 . \label{inmub}
\end{equation}

Dealing with two dimensions, we have
$\bar{\beta}_{2}=0.5-\lambda+\lambda^{2}$ and
$\tr(\bro^{2})=1-2\lambda+2\lambda^{2}=2\bar{\beta}_{2}$, where
$\lambda$ is the minimal eigenvalue of $\bro$. Hence, the right-hand
side of (\ref{eqren1mub}) is equal to the right-hand side of
(\ref{inmub}). In other words, the result (\ref{eqren1mub})
reproduces the previous one from \cite{rastmubs}. Note that both the inequalities
(\ref{eqren1mub}) and (\ref{inmub}) are based on an estimation of
the maximal probability from above. To compare these inequalities
with (\ref{guninub}), we plot the lower bounds on the average
min-entropy in Fig. \ref{fig2}. For a unification with other
pictures, the curve
$\bar{\beta}_{2}\mapsto-\ln\bigl(\widetilde{\Upsilon}_{1}^{(3)}(\bar{\beta}_{2})\bigr)$
is shown as well. In effect, the picture is generally similar to
what is seen in Fig. \ref{fig1}. Distinctions between estimates are
maximal on the left point, when $\bar{\beta}_{2}$ takes its minimal
acceptable value. For very mixed states, the inequalities
(\ref{eqren1mub}) and (\ref{inmub}) exceed the right-hand side of
(\ref{guninub}) almost $1.5$ times. For pure states, our curves
improve the right-hand side of (\ref{guninub}) by values of order
$3$ \%.

\begin{figure*}
\centering \includegraphics[height=7.6cm]{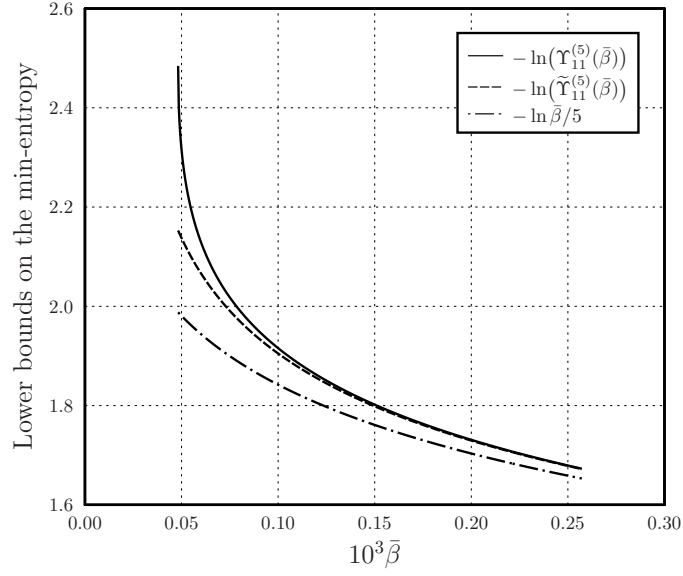}
\caption{\label{fig3} The three lower bounds on the min-entropy
versus the parameter (\ref{betak}) for the $5$-design with $12$
vertices. For convenience of presentation, the abscissa shows
$10^{3}\bar{\beta}$.}
\end{figure*}

The following example is the $5$-design with $K=12$ vertices forming
an icosahedron. The quantity $\Upsilon_{11}^{(5)}(\bar{\beta})$ was
calculated by means of the Newton--Raphson method. In Fig.
\ref{fig3}, we plot the bounds (\ref{gunin}), (\ref{eqres1}), and
(\ref{eqres11}) for this example. Like the example with octahedron, the
improvement due to both the formulas (\ref{eqres1}) and
(\ref{eqres11}) turns out to be significant. In effect,
the difference between (\ref{eqres1}) and (\ref{eqres11})
is maximal for sufficiently mixed states. For pure states, these
results enhance the right-hand side of (\ref{gunin}) by values of
order $1$ \%. For small values of $\bar{\beta}$, the right-hand
side of (\ref{eqres1}) exceeds the lower entropic bound
(\ref{gunin}) approximately $1.25$ times. In general, reached
improvements are considerable. Even in the state-independent
formulation, they are of interest.

\begin{figure*}
\centering \includegraphics[height=7.6cm]{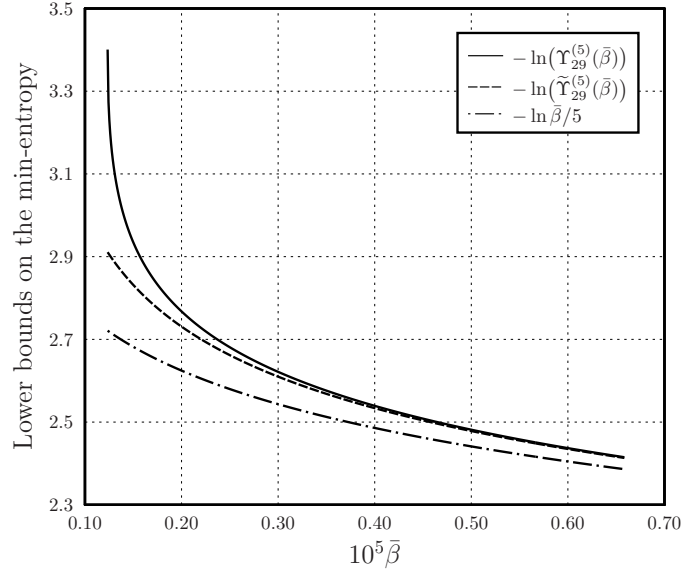}
\caption{\label{fig4} The three lower bounds on the min-entropy
versus the parameter (\ref{betak}) for the $5$-design with $30$
vertices. For convenience of presentation, the abscissa shows
$10^{5}\bar{\beta}$.}
\end{figure*}

There is also the $5$-design with $K=30$ vertices forming an
icosidodecahedron. On the average, actual values of probabilities
becomes lesser in view of increased number of outcomes. In addition,
acceptable values of the parameter (\ref{betak}) are lesser than in
the previous example. In Fig. \ref{fig4}, we plot the bounds
(\ref{gunin}), (\ref{eqres1}), and (\ref{eqres11}) for the example
with icosidodekahedron. The general picture is quite similar to what
is seen on Fig. \ref{fig3}. When $\bar{\beta}$ is close to its
maximal acceptable value, the difference between (\ref{eqres1}) and
(\ref{gunin}) is relatively small, whereas the lower bounds
(\ref{eqres1}) and (\ref{eqres11}) become coinciding. For small
values of $\bar{\beta}$, distinctions between the three bounds are
significant too.  Overall, this example also characterizes a
performance of the uncertainty relations (\ref{eqres1}) and
(\ref{eqres11}).

We have compared the uncertainty relations (\ref{gunin}),
(\ref{eqres1}), and (\ref{eqres11}) within the three examples of
quantum designs in two dimensions. It turned out that both the
results (\ref{eqres1}) and (\ref{eqres11}) allow us to reach
improved bounds. In the state-independent formulation, there is no
essential distinctions between the results (\ref{eqres1}) and
(\ref{eqres11}). When we know
$\tr\bigl(\bro^{\otimes{t}}\psymt\bigr)$ exactly or approximately,
the uncertainty relation (\ref{eqres1}) is clearly preferable. The
difference between the new bounds and (\ref{gunin}) is minimal for
pure states, when $\bar{\beta}=K^{1-t}d^{\,t}\,\cald_{d}^{(t)}$. For
$d=2$, this difference is comparatively small. It can be estimated
due to the formulation (\ref{eqres11}). Indeed, we have
\begin{align}
-\ln\bigl(\widetilde{\Upsilon}_{K-1}^{(t)}(\bar{\beta})\bigr)+\frac{1}{t}\,\ln\bar{\beta}=
-\ln\bigl(1-\chi_{K-1}^{(t)}(\bar{\beta})\bigr)\geq\chi_{K-1}^{(t)}(\bar{\beta})
\, , \label{minbo1}\\
\chi_{K-1}^{(t)}(\bar{\beta})=
\frac{\bar{\beta}^{\,-1/t}\left(1-\bar{\beta}^{\,1/t}\right)^{t}}{t(K-1)^{t-1}\bar{\beta}^{\,1-1/t}-t\bigl(1-\bar{\beta}^{\,1/t}\bigr)^{t-1}}
\ . \nonumber
\end{align}
Substituting $\bar{\beta}=K^{1-t}d^{\,t}\,\cald_{d}^{(t)}$, we can
estimate the mentioned difference. Overall, the quantity
$\chi_{K-1}^{(t)}(\bar{\beta})$ {\it per se} provides sufficiently precise
characterization. The result (\ref{minbo1}) gives a simple tool of comparing the two
state-independent formulations for quantum designs with other values
of $d$ and $K$.

The derived uncertainty relations directly lead to entropic steering
inequalities. The phenomenon of steering initially noticed by
Schr\"{o}dinger was rigorously formalized in \cite{wisem07}. For a
general discussion of this important issue, see the review
\cite{ucno20} and references therein. The writers of \cite{brun18} focused
on steering inequalities involving two entropies, but an extension
to more items is straightforward \cite{guhne19}. The
following conditions were formulated in \cite{brun18}. First, the considered
entropic uncertainty relations should be true, when they are conditioned on
a classical random variable. The respective argument in
application to R\'{e}nyi entropies was presented in \cite{brun18}.
Second, the considered entropies should be non-increasing under
conditioning on additional information. In view of the first
restriction, we have to use the state-independent formulation of
uncertainty relations.

To convert (\ref{eqres2n}) into entropic steering inequality, a
conditional form of the R\'{e}nyi entropy should be incorporated.
There is no generally accepted definition of conditional R\'{e}nyi
entropy \cite{tma12,fehr14}. For a non-negative real order $\alpha$, Arimoto
\cite{arim77} suggested the definition
\begin{equation}
R_{\alpha}(X|Z):=\frac{\alpha}{1-\alpha}
\,\ln\left\{\sum\nolimits_{z}p(z)\left(\sum\nolimits_{x}p(x|z)^{\alpha}\right)^{1/\alpha}\right\}
 . \label{crea}
\end{equation}
The corresponding conditional min-entropy is written as \cite{iws14}
\begin{equation}
R_{\infty}(X|Z):=
-\ln\!\left(\sum\nolimits_{z}p(z)\ \underset{x}{\max}\,p(x|z)\right)
 . \label{crein}
\end{equation}
Properties of these entropies with some applications are discussed
in \cite{fehr14,iws14}. In particular, the above conditional entropy
cannot increase under conditioning on additional information. This
property was originally proved by Arimoto \cite{arim77} and
reconsidered later in \cite{fehr14}. So, the entropies (\ref{crea})
and (\ref{crein}) obey the second condition formulated in
\cite{brun18}.

Suppose that Alice and Bob share a bipartite quantum state
$\bro_{AB}$, and repeat this any number of times. Alice performs on
her subsystem a measurement chosen from the set of POVMs
$\bigl\{\clf^{(m)}\bigr\}_{m=1}^{M}$. So, the actual state of Bob's
subsystem is conditioned on Alice's result. Bob's conditioned state
is subjected to a measurement chosen accordingly from the set
$\bigl\{\cle^{(m)}\bigr\}_{m=1}^{M}$. We consider measurements
assigned to a quantum $t$-design. Using the generated probabilities
and classical side information from Alice, Bob takes the conditional
$\alpha$-entropies $R_{\alpha}\bigl(\cle^{(m)}|\clf^{(m)}\bigr)$
according to (\ref{crea}). The latter depends on a shared state,
though we do not mark this dependence explicitly. Following the
approach of \cite{guhne19,brun18}, the inequality (\ref{eqres2in})
leads to
\begin{equation}
\frac{1}{M}\>\sum_{m=1}^{M}R_{\alpha}\bigl(\cle^{(m)}|\clf^{(m)}\bigr)
\geq-\,\frac{\alpha-t}{\alpha-1}\,\ln\bigl(\Upsilon_{n-1}^{(t)}(n^{1-t}d^{\,t}\,\cald_{d}^{(t)})\bigr)
-\frac{\ln(n^{1-t}d^{\,t}\,\cald_{d}^{(t)})}{\alpha-1}
\ , \label{stes2n}
\end{equation}
where $\alpha\geq{t}$. The entropic steering inequality
(\ref{stes2n}) allows us to enhance somehow the steering inequalities
given in \cite{guhne19}. Due to the
state-independent form of (\ref{lpmax}), we have another steering inequality
\begin{equation}
\frac{1}{M}\>\sum_{m=1}^{M}\sum_{\ell}\,p_{\ell}(\clf^{(m)};\bro_{A})\ \underset{j}{\max}\,p_{j}(\cle^{(m)};\bro_{B\ell}^{(m)})
\leq\Upsilon_{n-1}^{(t)}(n^{1-t}d^{\,t}\,\cald_{d}^{(t)})
\, , \label{stmax}
\end{equation}
where the reduced densities are defined by
$\bro_{A}=\tr_{B}(\bro_{AB})$ and
\begin{equation}
p_{\ell}(\clf^{(m)};\bro_{A})\>\bro_{B\ell}^{(m)}=\tr_{A}\bigl\{(\mfe_{\ell}^{(m)}\otimes\pen_{d})^{1/2}\bro_{AB}(\mfe_{\ell}^{(m)}\otimes\pen_{d})^{1/2}\bigr\}
\, . \nonumber
\end{equation}
In principle, relations in terms of maximal probabilities may be
more suitable in some questions. Applications of (\ref{lpmax}) and (\ref{lpmak}) to
posing entanglement and steerability criteria could be the subject
of separate research.

\section{Conclusions}\label{sec5}

New uncertainty relations in terms of R\'{e}nyi entropies were
derived for POVM measurements assigned to a quantum design. The
method used is a natural extension of the argument proposed in
\cite{rastmubs}. The principal point is to estimate from above the
average maximal probability. It is possible due to the fact that the
sum of powers $t$ of rescaled probabilities is calculated exactly
for a quantum $t$-design. The corresponding answer is expressed via
trace of the projector onto the symmetric subspace of $t$-fold
tensor-product space. Using done estimate of the average maximal
probability, we immediately obtain uncertainty relations in terms of
min-entropies as well as relations of the Landau--Pollak type.
Moreover, one has easily obtained lower bounds on the average
$\alpha$-entropy for arbitrary $\alpha\geq{t}$.

The presented uncertainty relations are expressed in terms of
specific root of algebraic equation of power $t$. Except for simple
cases, this root cannot be expressed analytically using radicals.
Nevertheless, the Newton--Raphson method is quite applicable in all
particular situations. On the other hand, the first Newton--Raphson
step leads to the explicit bound written analytically. Moreover,
such an estimation is sufficiently good in the state-independent
formulation. It must be stressed that for the maximally mixed state
our lower bound on the min-entropy is saturated. A utility of the
new bounds was demonstrated within several examples of quantum
designs in two dimensions. Also, we briefly discussed steering
inequalities based on the derived uncertainty relations.

\appendix

\section{Some properties of roots of the interest}\label{roots}

This appendix aims to show that, for fixed $t$ and $n$, the quantity
$\Upsilon_{n-1}^{(t)}(\beta)$ is increasing and concave with respect
to $\beta$. We restrict our consideration to the case $t\geq2$ and
$n\geq2$. Let us consider the equation
\begin{equation}
(n-1)^{t-1}y^{t}+(1-y)^{t}=(n-1)^{t-1}\beta
\, . \label{xttdb}
\end{equation}
The quantity $\Upsilon_{n-1}^{(t)}(\beta)$ is defined as the maximal
real root of this equation. Taking the derivative of both the sides
of (\ref{xttdb}) with respect to $\beta$ gives
\begin{equation}
t\>\frac{\partial{y}}{\partial\beta}\left[(n-1)^{t-1}y^{t-1}-(1-y)^{t-1}\right]=(n-1)^{t-1}
\, . \label{xttdb1}
\end{equation}
The equality $(n-1)^{t-1}y^{t-1}=(1-y)^{t-1}$ implies $y=1/n$,
whence $(n-1)^{t-1}y^{t-1}>(1-y)^{t-1}$ for $y>1/n$. Combining
this with (\ref{xttdb1}) results in the inequality
$\partial{y}/\partial\beta>0$. For the
second derivative, we get
\begin{equation}
t\>\frac{\partial^{2}y}{\partial\beta^{2}}\left[(n-1)^{t-1}y^{t-1}-(1-y)^{t-1}\right]
+t(t-1)\Bigl(\frac{\partial{y}}{\partial\beta}\Bigr)^{2}\left[(n-1)^{t-1}y^{t-2}+(1-y)^{t-2}\right]
=0
\, . \label{xttdb2}
\end{equation}
For the considered ranges of variables, we see the following. The
multiplier of the second derivative and the second summand in the
left-hand side of (\ref{xttdb2}) are both non-negative. Hence, we
finally obtain $\partial^{2}y/\partial\beta^{2}\leq0$ as claimed.

\section{Exact expressions for the cubic case}\label{cas34}

In this section, we present auxiliary formulas for the case $t=3$.
According to the standard treatment, the cubic equation of interest
is transformed to the reduced form
\begin{equation}
\xi^{3}+p\,\xi+q=0
\, , \qquad
\Upsilon_{n-1}^{(3)}(\beta)=\xi-\frac{1}{n^{2}-2n}
\ . \label{redce}
\end{equation}
In our case, the coefficients are written as
\begin{equation}
p=-\,\frac{3(n-1)^2}{(n^{2}-2n)^{2}}
\   , \qquad
q=\frac{3n^{2}-6n+2}{(n^{2}-2n)^{3}}+\frac{1-(n-1)^{2}\beta}{n^{2}-2n}
\ . \label{redce1}
\end{equation}
The real root of the equation (\ref{redce}) is expressed in line
with Cardano's formula (see, e.g., section 1.8-3 of the handbook
\cite{korn2000}). Finally, the desired answer is posed by
\begin{equation}
\Upsilon_{n-1}^{(3)}(\beta)=\sqrt[3]{-\,\frac{q}{2}+\sqrt{Q}\,}+\sqrt[3]{-\,\frac{q}{2}-\sqrt{Q}\,}-\frac{1}{n^{2}-2n}
\ , \qquad
Q=\biggl(\frac{p}{3}\biggr)^{\!3}+\biggl(\frac{q}{2}\biggr)^{\!2}
 . \label{redce3}
\end{equation}
In the case considered, one should use the principal cube root given
that the argument of complex numbers is taken between $-\pi$ and
$+\pi$. The cubic equation has three real roots of which at least
two are equal, or three different real roots, if $Q$ is zero or
negative, respectively \cite{korn2000}. The latter is in agreement
with the notes given right after (\ref{asymp}). The above
expressions are not suitable for $n=2$. For $n=2$ and odd $t$, the
equation (\ref{xttdb}) reduces to the power $t-1$. Substituting
$n=2$ and $t=3$ into (\ref{xttdb}) leads to the equation
$3y^{2}-3y+1=\beta$, whence
\begin{equation}
\Upsilon_{1}^{(3)}(\beta)=\frac{1}{2}+\sqrt{\frac{4\beta-1}{12}}
\ . \label{upsn2}
\end{equation}
In a similar manner, one can treat the quartic case. There exist
several formulations, mainly due to Ferrari, Descartes and Euler.
Explicit easy-to-handle expressions using radicals are presented in
\cite{yacf12}. We refrain from reproducing the details here.


\begin{thebibliography}{00}

\bibitem{heisenberg}%---------------------------------------------
Heisenberg W 1927 {\it Z. Phys.} {\bf 43} 172 

\bibitem{kennard}%------------------------------------------------
Kennard E H 1927 {\it Z. Phys.} {\bf 44} 326 

\bibitem{robert}%--------------------------------------------------
Robertson H P 1929 {\it Phys. Rev.} {\bf 34} 163

\bibitem{lahti}%-----------------------------------------------
Busch P, Heinonen T and Lahti P J 2007 {\it Phys. Rep.} {\bf 452} 155 

\bibitem{gour16}%------------------------------------------------
Narasimhachar V, Poostindouz A and Gour G 2016 {\it New J. Phys.} {\bf 18} 033019 

\bibitem{mdsrin03}%------------------------------------------------
Srinivas M D 2003 {\it Pramana} {\bf 60} 1137 

\bibitem{paban13}%--------------------------------------------------
Distler J and Paban S 2013 {\it Phys. Rev. A} {\bf 87} 062112 

\bibitem{rozp17}%--------------------------------------------------
Rozp\c{e}dek F, Kaniewski J, Coles P J and Wehner S 2017 {\it New J. Phys.} {\bf 19} 023038 

\bibitem{deutsch}%---------------------------------------------------
Deutsch D 1983 {\it Phys. Rev. Lett.} {\bf 50} 631 

\bibitem{maass}%----------------------------------------------------
Maassen H and Uffink J B M 1988 {\it Phys. Rev. Lett.} {\bf 60} 1103 

\bibitem{huang12}%------------------------------------------------
Huang Y 2012 {\it Phys. Rev. A} {\bf 86} 024101 

\bibitem{macpa14}%------------------------------------------------
Maccone L and Pati A K 2014 {\it Phys. Rev. Lett.} {\bf 113} 260401 

\bibitem{ww10}%-------------------------------------------------
Wehner S and Winter A 2010 {\it New J. Phys.} {\bf 12} 025009 

\bibitem{brud11}%------------------------------------------------
Bia{\l}ynicki-Birula I and Rudnicki {\L} 2011 Entropic uncertainty relations in quantum physics {\it Statistical Complexity} (Berlin: Springer) pp 1--34

\bibitem{cbtw17}%------------------------------------------------
Coles P J, Berta M, Tomamichel M and Wehner S 2017 {\it Rev. Mod. Phys.} {\bf 89} 015002 

\bibitem{hall18}%------------------------------------------------
Hall M J W 2018 {\it J. Phys. A: Math. Theor.} {\bf 51}, 364001 

\bibitem{cerf19}%------------------------------------------------
Hertz A and Cerf N J 2019 {\it J. Phys. A: Math. Theor.} {\bf 52} 173001 

\bibitem{bccrr10}%------------------------------------------
Berta M, Christandl M, Colbeck R, Renes J M and Renner R 2010 {\it Nature Phys.} {\bf 6} 659 

\bibitem{cp2014}%-------------------------------------------------
Coles PJ and Piani M 2014 {\it Phys. Rev. A} {\bf 89} 022112 

\bibitem{bww16}%--------------------------------------------------
Berta M, Wehner S and Wilde M M 2016 {\it New J. Phys.} {\bf 18} 073004 

\bibitem{ccyz12}%------------------------------------------
Coles P J, Colbeck R, Yu L and Zwolak M 2012 {\it Phys. Rev. Lett.} {\bf 108} 210405 

\bibitem{fgg2013}%-------------------------------------------------
Friedland S, Gheorghiu V and Gour G 2013 {\it Phys. Rev. Lett.} {\bf 111} 230401 

\bibitem{prz2013}%-------------------------------------------------
Pucha{\l}a Z, Rudnicki {\L} and \.Zyczkowski K 2013 {\it J. Phys. A: Math. Theor.} {\bf 46} 272002 

\bibitem{rpz2014}%-------------------------------------------------
Rudnicki {\L}, Pucha{\l}a Z and \.Zyczkowski K 2014 {\it Phys. Rev. A} {\bf 89} 052115 

\bibitem{rz2016}%-------------------------------------------------
Rastegin A E and \.{Z}yczkowski K 2016 {\it J. Phys. A: Math. Theor.} {\bf 49} 355301 

\bibitem{zbig2018}%-------------------------------------------------
Pucha{\l}a Z, Rudnicki {\L}, Krawiec A and \.{Z}yczkowski K 2018 {\it J. Phys. A: Math. Theor.} {\bf 51} 175306 

\bibitem{zozor2013}%-------------------------------------------------
Zozor S, Bosyk G M and Portesi M 2013 {\it J. Phys. A: Math. Theor.} {\bf 46} 465301 

\bibitem{zozor2014}%-------------------------------------------------
Zozor S, Bosyk G M and Portesi M 2014 {\it J. Phys. A: Math. Theor.} {\bf 47} 495302 

\bibitem{baek14}%------------------------------------------------
Baek K, Farrow T and Son W 2014 {\it Phys. Rev. A} {\bf 89} 032108 

\bibitem{zzy15}%------------------------------------------------
Zhang J, Zhang Y and Yu C-S 2015 {\it Quantum Inf. Process.} {\bf 14} 2239 

\bibitem{rastann}%------------------------------------------------
Rastegin A E 2016 {\it Ann. Phys. (Berlin)} {\bf 528} 835 

\bibitem{rastent}%------------------------------------------------
Rastegin A E 2018 {\it Entropy} {\bf 20} 354 

\bibitem{bz10}%--------------------------------------------------------------------------
Durt T, Englert B-G, Bengtsson I and \.{Z}yczkowski K 2010 {\it Int. J. Quantum Inf.} {\bf 8} 535 

\bibitem{rbksc04}%------------------------------------------------
Renes J, Blume-Kohout R, Scott A and Caves C 2004 {\it J. Math. Phys.} {\bf 45} 2171 

\bibitem{scott10}%------------------------------------------------
Scott A J and Grassl M 2010 {\it J. Math. Phys.} {\bf 51} 042203 

\bibitem{scott17}%------------------------------------------------
Scott A J 2017 SICs: Extending the list of solutions arXiv:1703.03993 [quant-ph] 

\bibitem{abfg19}%------------------------------------------------
Appleby M, Bengtsson I, Flammia S and Goyeneche D 2019 {\it J. Phys. A: Math. Theor.} {\bf 52} 295301 

\bibitem{appleby05}%------------------------------------------------
Appleby D M 2005 {\it J. Math. Phys.} {\bf 46} 052107 
 
\bibitem{fuchs17}%------------------------------------------------
Fuchs C A, Hoang M C and Stacey B C 2017 {\it Axioms} {\bf 6} 21 

\bibitem{af17}%------------------------------------------------
Appleby M, Flammia S, McConnell G and Yard J 2017 {\it Found. Phys.} {\bf 47} 1042 

\bibitem{ab19}%------------------------------------------------
Appleby M and Bengtsson I 2019 {\it J. Math. Phys.} {\bf 60} 062203 

\bibitem{delsart77}%------------------------------------------------
Delsarte P, Goethals J and Seidel J 1977 {\it Geom. Dedicata} {\bf 6} 363 

\bibitem{scottjpa}%------------------------------------------------
Scott A J 2006 {\it J. Phys. A: Math. Gen.} {\bf 39} 13507 

\bibitem{ambain07}%------------------------------------------------
Ambainis A and Emerson J 2007 Quantum $t$-designs: $t$-wise independence in the quantum world arXiv:quant-ph/0701126 

\bibitem{hoggar82}%-------------------------------------------------
Hoggar S G 1982 {\it Eur. J. Combin.} {\bf 3} 233 

\bibitem{scottjpa8}%------------------------------------------------
Scott A J 2008 {\it J. Phys. A: Math. Theor.} {\bf 41} 055308 

\bibitem{scott9}%------------------------------------------------
Roy A and Scott A J 2009 {\it Des. Codes Cryptogr.} {\bf 53} 13 

\bibitem{dcel9}%------------------------------------------------
Dankert C, Cleve R, Emerson J and Livine E 2009 {\it Phys. Rev. A} {\bf 80} 012304 

\bibitem{bhmn19}%------------------------------------------------
Bae J, Hiesmayr B C and McNulty D 2019 {\it New J. Phys.} {\bf 21} 013012 

\bibitem{kwg19}%------------------------------------------------
Ketterer A, Wyderka N and G\"{u}hne O 2019 {\it Phys. Rev. Lett.} {\bf 122} 120505

\bibitem{cggz19}%------------------------------------------------
Czartowski J, Goyeneche D, Grassl M and \.{Z}yczkowski K 2019 {\it Phys. Rev. Lett.} {\bf 124} 090503 

\bibitem{guhne19}%------------------------------------------------
Ketterer A and G\"{u}hne O 2020 {\it Phys. Rev. Research} {\bf 2} 023130 

\bibitem{rastmubs}%------------------------------------------------
Rastegin A E 2013 {\it Eur. Phys. J. D} {\bf 67} 269 

\bibitem{cons98}%------------------------------------------------
Conway J H and Sloane N J A 1998 {\it Sphere Packing, Lattices and Groups} (New York: Springer-Verlag) 

\bibitem{bengtsson}%------------------------------------------------
Bengtsson I and \.{Z}yczkowski K 2017 {\it Geometry of Quantum States: An Introduction to Quantum Entanglement} (Cambridge: Cambridge University Press)

\bibitem{seym1984}%------------------------------------------------
Seymour P D and Zaslavsky T 1984 {\it Adv. Math.} {\bf 52} 213 

\bibitem{hardin96}%------------------------------------------------
Hardin R H and Sloane N J A 1996 {\it Discrete Comput. Geom.} {\bf 15} 429 

\bibitem{gross07}%------------------------------------------------
Gross D, Audenaert K and Eisert J 2007 {\it J. Math. Phys.} {\bf 48} 052104 

\bibitem{renyi61}%-------------------------------------------------
R\'{e}nyi A 1961 On measures of entropy and information {\it Proceedings of the 4th Berkeley Symposium on Mathematical Statistics and Probability} (Berkeley, CA: University of California Press) pp 547--61

\bibitem{Kam98}%------------------------------------------------
Kamimura R 1998 {\it Algorithmica} {\bf 22} 173 

\bibitem{ja04}%-------------------------------------------
Jizba P and Arimitsu T 2004 {\it Ann. Phys.} {\bf 312} 17 

\bibitem{rastkyb}%------------------------------------------------
Rastegin A E 2012 {\it Kybernetika} {\bf 48} 242 

\bibitem{rastrai}%------------------------------------------------
Rastegin A E 2015 {\it RAIRO--Theor. Inf. Appl.} {\bf 49} 67 

\bibitem{rastineq}%------------------------------------------------
Rastegin A E 2019 {\it Quantum Inf. Process.} {\bf 18} 276 

\bibitem{rastcomb}%------------------------------------------------
Rastegin A E 2016 {\it Graphs Combin.} {\bf 32} 2625 

\bibitem{MWB10}%---------------------------------------------------
Mandayam P, Wehner S and Balachandran N 2010 {\it J. Math. Phys.} {\bf 51} 082201 

\bibitem{ngbw12}%-------------------------------------------------
Ng H Y N, Berta M and Wehner S 2012 {\it Phys. Rev. A} {\bf 86} 042315 

\bibitem{cirac18}%-------------------------------------------------
Vermersch B, Elben A, Dalmonte M, Cirac J I and Zoller P 2018 {\it Phys. Rev. A} {\bf 97} 023604 

\bibitem{pollak61}%-------------------------------------------------
Landau H J and Pollak H O 1961 {\it Bell Syst. Tech. J.} {\bf 40} 65 

\bibitem{immy2007}%-------------------------------------------------
Miyadera T and Imai H 2007 {\it Phys. Rev. A} {\bf 76} 062108 

\bibitem{bosyk2014}%-------------------------------------------------
Bosyk G M, Zozor S, Portesi M, Os\'{a}n T M and Lamberti P W 2014 {\it Phys. Rev. A} {\bf 90} 052114

\bibitem{vicen05}%------------------------------------------------
de Vicente J I and S\'{a}nchez-Ruiz J 2005 {\it Phys. Rev. A} {\bf 71} 052325 

\bibitem{rastosid}%------------------------------------------------
Rastegin A E 2015 {\it Open Syst. Inf. Dyn.} {\bf 22} 1550005 

\bibitem{wisem07}%-------------------------------------------------
Wiseman H M, Jones S J and Doherty A C 2007 {\it Phys. Rev. Lett.} {\bf 98} 140402 

\bibitem{ucno20}%-------------------------------------------------
Uola R, Costa A C S, Nguyen H C and G\"{u}hne O 2020 {\it Rev. Mod. Phys.} {\bf 92} 15001 

\bibitem{brun18}%-------------------------------------------------
Kriv\'{a}chy T, Fr\"{o}wis F and Brunner N 2018 {\it Phys. Rev A} {\bf 98} 062111 

\bibitem{tma12}%-------------------------------------------------
Teixeira A, Matos A and Antunes L 2012 {\it IEEE Trans. Inf. Theory} {\bf 58} 4273 

\bibitem{fehr14}%-------------------------------------------------
Fehr S and Berens S 2014 {\it IEEE Trans. Inf. Theory} {\bf 60} 6801 

\bibitem{arim77}%-------------------------------------------------
Arimoto S 1977 Information measures and capacity of order $\alpha$ for discrete memoryless channels {\it Topics in Information Theory} ({\it Colloquia Mathematica Societatis J\'{a}nos Bolyai} Vol 16)  (Amsterdam: North-Holland) pp 41--52

\bibitem{iws14}%-------------------------------------------------
Iwamoto M and Shikata J 2014 Secret sharing schemes based on min-entropies {\it 2014 IEEE Int. Symp. on Information Theory} pp 401--05

\bibitem{korn2000}%------------------------------------------------
Korn G A and Korn T M 2000 {\it Mathematical Handbook for Scientists and Engineers} (New York: Dover) 

\bibitem{yacf12}%------------------------------------------------
Yacoub M D and Fraidenraich G 2012 {\it Math. Gaz.} {\bf 96} 271 


\end{thebibliography}
\end{document}